\documentclass[twocolumn,nodate,showpacs,preprintnumbers,amsmath,amssymb,aps,prd]{revtex4}
\usepackage{setspace}
\usepackage{amsmath,amssymb,amsfonts,amsthm,latexsym}
\usepackage{graphicx}
\usepackage{enumerate}
\usepackage{colordvi}

\newcommand*{\Z}{{\mathbb Z}}

\allowdisplaybreaks
\begin{document}

\title{Alternative quantization of the Hamiltonian in isotropic loop quantum cosmology}
\author{Jinsong Yang${}^1$}
\author{You Ding${}^{1,2}$}
\author{Yongge Ma${}^{1}$}
\email{mayg@bnu.edu.cn} \affiliation{\small{${}^1$Department of
Physics, Beijing Normal
University, Beijing 100875, China}\\
\small{${}^{2}$ Centre de Physique Th\'eorique de Luminy,
Universit\'e de la M\'editerran\'ee, F-13288 Marseille, EU}}

\begin{abstract}
Since there are quantization ambiguities in constructing the
Hamiltonian constraint operator in isotropic loop quantum
cosmology, it is crucial to check whether the key features of loop
quantum cosmology, such as the quantum bounce and effective
scenario, are robust against the ambiguities. In this paper, we
consider a typical quantization ambiguity arising from the
quantization of the field strength of the gravitational
connection. An alternative Hamiltonian constraint operator is
constructed, which is shown to have the correct classical limit by
the semiclassical analysis. The effective Hamiltonian
incorporating higher order quantum corrections is also obtained.
In the spatially flat FRW model with a massless scalar field, the
classical big bang is again replaced by a quantum bounce.
Moreover, there are still great possibilities for the expanding
universe to recollapse due to the quantum gravity effect. Thus,
these key features are robust against this quantization ambiguity.
\end{abstract}
\maketitle

\section{Introduction}

An important motivation of the theoretical search for a quantum
theory of gravity is the expectation that the singularities
predicted by classical general relativity would be resolved by the
quantum gravity theory. This expectation has been confirmed by the
recent study of certain isotropic models in loop quantum cosmology
(LQC) \cite{bojowald,apsL,aps}, which is a simplified
symmetry-reduced model of a full background-independent quantum
theory of gravity \cite{lrr}, known as loop quantum gravity (LQG)
\cite{rev,rov,thie,hhm}. In loop quantum cosmological scenario for
a universe filled with a massless scalar field, the classical
singularity gets replaced by a quantum bounce
\cite{aps,bojow-r,AshR}. Moreover, it is revealed in the effective
scenarios that there are great possibilities for a spatially flat
FRW expanding universe to recollapse due to the quantum gravity
effect \cite{dmy}. However, as in the ordinary quantization
procedure, there are quantization ambiguities in constructing the
Hamiltonian constraint operator. Thus a  crucial question arises.
Whether the above significant results come from certain particular
treatment of quantization? Before confirming the robustness of the
key results against the quantization ambiguities, one could not
believe that they are not some artifact under particular
assumptions.

In this paper, we consider a typical quantization ambiguity
arising from the quantization of the field strength of the
gravitational connection. An alternative Hamiltonian constraint
operator is constructed. Semiclassical states are then employed to
show that the new Hamiltonian operator has the correct classical
limit. The effective Hamiltonian incorporating higher order
quantum corrections is also obtained. In the spatially flat FRW
model with a massless scalar field, the classical big bang is
again replaced by a quantum bounce. Moreover, there are still
great possibilities for the expanding universe to recollapse due
to the quantum gravity effect.

In the spatially flat isotropic model of LQC, one has to first
introduce an elementary cell ${\cal V}$ and restrict all
integrations to this cell. Fix a fiducial flat metric
${{}^o\!q}_{ab}$ and denote by $V_o$ the volume of the elementary
cell ${\cal V}$ in this geometry. The gravitational phase space
variables
---the connections $A_a^i$ and the density-weighted triads
$E^a_i$--- can be expressed as
\begin{align}
A_a^i = c\, V_o^{-1/3}\,\, {}^o\!\omega_a^i \quad\mathrm{and}\quad
E^a_i = p\, V_o^{-2/3}\,\sqrt{{}^o\!q}\,\, {}^o\!e^a_i,
\end{align} where
$({}^o\!\omega_a^i, {}^o\!e^a_i)$ are a set of orthonormal
co-triads and triads compatible with ${{}^o\!q}_{ab}$ and adapted
to the edges of the elementary cell ${\cal V}$. The basic
(nonvanishing) Poisson bracket is given by
\begin{align}
\label{bracket} \{c,\, p\} = \frac{8\pi G\gamma}{3},
\end{align}
where $G$ is the Newton's constant and $\gamma$ is the
Barbero-Immirzi parameter.

To pass to the quantum theory, one constructs a kinematical Hilbert
space ${\cal
H}^{\mathrm{grav}}_{\mathrm{kin}}=L^2(\mathbb{R}_{\mathrm{Bohr}},{\mathrm{d}}\mu_{\mathrm{Bohr}})$,
where $\mathbb{R}_{\mathrm{Bohr}}$ is the Bohr compactification of
the real line and ${\mathrm{d}}\mu_{\mathrm{Bohr}}$ is the Haar
measure on it \cite{math}. The abstract $*$-algebra represented on
the Hilbert space is based on the holonomies of connection $A^i_a$.
In the Hamiltonian constraint of LQG, the gravitational connection
$A^i_a$ appears through its curvature $F^i_{ab}$. Since there exists
no operator corresponding to $c$, only holonomy operators are well
defined. Hence one is led to express the curvature in terms of
holonomies. Similarly, in the improved dynamics setting of LQC
\cite{aps}, to express the curvature one employed the holonomies
\begin{align}
\label{hol} h_i^{(\bar{\mu})}:=\cos\frac{\bar{\mu}
c}{2}\,\mathbb{I}+2\sin\frac{\bar{\mu} c}{2}\,\tau_i
\end{align}
along an edge parallel to the triad ${}^o\!e^a_i$ of length
$\bar{\mu}\sqrt{|p|}$ with respect to the physical metric
$q_{ab}$, where $\mathbb{I}$ is the identity $2\times2$ matrix and
$\tau_i=-i\sigma_i/2$ ($\sigma_i$ are the Pauli matrices). Thus,
the elementary variables could be taken as the functions
$\exp(i\bar{\mu} c/2)$ and the physical volume $V=|p|^{3/2}$ of
the cell, which have unambiguous operator analogs.

\section{The alternative Hamiltonian constraint operator}

In general relativity, the dynamics of a gravitational system is
determined by the Hamiltonian constraint. Many problems, such as
the big-bang singulary, arise in classical dynamics. One expects
that some quantum dynamics can resolve these problems. Hence, it
is important to have a well-defined Hamiltonian constraint
operator in LQC. An improved Hamiltonian constraint operator has
been constructed in \cite{aps}. However, there are quantization
ambiguities in the construction. In this section, we will
construct an alternative Hamiltonian constraint operator by a
different quantization procedure.

Because of spatial flatness and homogeneity, the gravitational
part of the Hamiltonian constraint of full general relativity is
simplified to the form
\begin{align}
\label{Cc} C_{\mathrm{grav}}=-\gamma^{-2}\int_{\cal
V}{\mathrm{d}}^3xN\epsilon_{ijk}F^i_{ab}e^{-1}E^{aj}E^{bk},
\end{align}
where $e:=\sqrt{|\det{E}|}$, the lapse $N$ is constant and we will
set it to one.

The procedure used in LQC (and specifically in \cite{aps,math}) so
far can be summarized as follows. The term involving the triads
can be written as
\begin{align}
\label{e}
\epsilon_{ijk}e^{-1}E^{aj}E^{bk}&=\sum_{k}\frac{\mathrm{sgn}(p)}{2\pi\gamma
G\bar{\mu} V_o^{1/3}}\,\,
    \epsilon^{abc}\,\,
    {}^o\!\omega_c^k\nonumber\\
    &\quad\times\mathrm{Tr}\left(h_k^{(\bar{\mu})}\left\{{h_k^{(\bar{\mu})}}^{-1},V\right\}\tau_i\right).
\end{align}
To express the curvature components $F^i_{ab}$ in terms of
holonomies, one considers a square $\Box_{ij}$ in the $i$-$j$ plane
spanned by a face of ${\cal V}$, each of whose sides has length
$\bar{\mu}\sqrt{|p|}$ with respect to $q_{ab}$. Then the $ab$
component of the curvature is given by
\begin{align}
\label{F1}
F^i_{ab}\tau_i=\lim_{Ar_\Box\rightarrow0}\frac{h_{\Box_{ij}}^{(\bar{\mu})}-1}{\bar{\mu}^2V_o^{2/3}}
{}^o\!\omega_{[a}^i{}^o\!\omega_{b]}^j,
\end{align}
where $Ar_\Box$ is the area of the square under consideration, and
the holonomy $h_{\Box_{ij}}^{(\bar{\mu})}$ around the square
$\Box_{ij}$ is just the product of holonomies along the four edges
of $\Box_{ij}$,
\begin{align}
h_{\Box_{ij}}^{(\bar{\mu})}=h_i^{(\bar{\mu})}h_j^{(\bar{\mu})}{h_i^{(\bar{\mu})}}^{-1}{h_j^{(\bar{\mu})}}^{-1}.
\end{align}
However, quantization ambiguities arise here, since the approach
to express the curvature components $F^i_{ab}$ in terms of
holonomies is not unique. Hence the corresponding operators in
different approaches will be different from each other. In the
following, we will consider an expression of the curvature
components different from Eq. (\ref{F1}). Taking account of the
definition (\ref{hol}) of holonomies, we have the identity
\begin{align}
\lim_{\bar{\mu}\rightarrow0}\frac{
h_i^{(\bar{\mu})}-{h_i^{(\bar{\mu})}}^{-1}}{\bar{\mu}}=\lim_{\bar{\mu}\rightarrow0}\frac{4\sin(\bar{\mu}
c/2)\,\tau_i}{\bar{\mu}}=2c\tau_i.
\end{align}
Hence the curvature of connection can be written in terms of the
holomomies as
\begin{align}
\label{F} F^k_{ab}\tau_k&=c^2V_o^{-2/3}\epsilon_{ijk}\,\,
{}^o\!\omega_a^i\,\, {}^o\!\omega_b^j\tau_k
\nonumber\\
&=\lim_{\bar{\mu}\rightarrow0}\frac{\left(h_i^{(\bar{\mu})}-{h_i^{(\bar{\mu})}}^{-1}\right)
\left(h_j^{(\bar{\mu})}-{h_j^{(\bar{\mu})}}^{-1}\right)}
{2\bar{\mu}^2V_o^{2/3}}\,\,
{}^o\!\omega_{[a}^i\,\, {}^o\!\omega_{b]}^j.
\end{align}
Combining Eqs. (\ref{e}) and (\ref{F}), the Hamiltonian constraint
can be written as
\begin{align}
C_{\mathrm{grav}}
&=-\lim_{\bar{\mu}\rightarrow0}\frac{\mathrm{sgn}(p)}{4\pi\gamma^3
    G\bar{\mu}^3
    }\epsilon^{ijk}\mathrm{Tr}\Big[\left(h_i^{(\bar{\mu})}-{h_i^{(\bar{\mu})}}^{-1}\right)\nonumber\\
&\quad\quad\quad\quad\times\left(h_j^{(\bar{\mu})}-{h_j^{(\bar{\mu})}}^{-1}\right)h_k^{(\bar{\mu})}
\left\{{h_k^{(\bar{\mu})}}^{-1},V\right\}\Big]\nonumber\\
&\equiv\lim_{\bar{\mu}\rightarrow0}{C^{\mathrm{R}}}^{(\bar{\mu})}_{\mathrm{grav}}.
\end{align}
Since the constraint is now expressed in terms of elementary
variables and their Poisson bracket, it can be promoted to a quantum
operator directly. The resulting alternative regulated constraint
operator with symmetric factor-ordering reads
\begin{align}
\hat{C^{\mathrm{R}}}^{(\bar{\mu})}_{\mathrm{grav}}&=\sin\frac{\bar{\mu}
    c}{2}\Big[\frac{12i\,\mathrm{sgn}(\hat{p})}{\pi\gamma^3
    \bar{\mu}^3\ell_p^2
    }\Big(\sin\frac{\bar{\mu}
    c}{2}\hat{V}\cos\frac{\bar{\mu}
    c}{2}\nonumber\\
    &\quad\quad\quad\quad-\cos\frac{\bar{\mu}
    c}{2}\hat{V}\sin\frac{\bar{\mu}
    c}{2}\Big)\Big]\sin\frac{\bar{\mu}
    c}{2},
\end{align}
where, for clarity, we have suppressed hats over the operators
$h_i^{(\bar{\mu})}$, $\sin(\bar{\mu} c/2)$ and $\cos(\bar{\mu}
c/2)$, and $\ell_p^2=\sqrt{G\hbar}$. To deal with the regulator
$\bar{\mu}$, we adopt the improved scheme \cite{aps}. We shrink the
length of holonomy edges, as measured by the physical metric
$q_{ab}$, to the value $\sqrt{\Delta}$, where
$\Delta=4\sqrt{3}\pi\gamma\ell^2_p$ is a minimum nonzero eigenvalue
of the area operator \cite{AshR}. Thus we are led to choose for
$\bar{\mu}$ a specific function $\bar{\mu}(p)$, given by
\begin{align}
\bar{\mu}=\sqrt{\Delta/|p|}\,.
\end{align}
It is convenient to work with the $v$-representation. In this
representation, states $|v\rangle$ constituting an orthonormal
basis in ${\cal H}^{\mathrm{grav}}_{\mathrm{kin}}$ is more
directly adapted to the volume operator $\hat{V}$,
\begin{align}
\hat{V}|v\rangle=\left(\frac{8\pi\gamma\ell_p^2}{6}\right)^{3/2}\frac{|v|}{K}|v\rangle,
\end{align}
where
\begin{align}
K=\frac{4}{3}\sqrt{\frac{\pi\gamma\ell_p^2}{3\Delta}}.
\end{align}
The action of
$\widehat{\exp(i\bar{\mu}c/2)}$ is given by
\begin{align}
\widehat{\exp(i\bar{\mu}c/2)}|v\rangle=|v+1\rangle.
\end{align}
Hence the alternative Hamiltonian constraint operator is given by
\begin{align}
\label{Cr}
\hat{C}^{\mathrm{R}}_{\mathrm{grav}}&=\sin\frac{\bar{\mu}
    c}{2}\Big[\frac{12i\,\mathrm{sgn}(v)}{\pi\gamma^3
    \bar{\mu}^3\ell_p^2
    }\Big(\sin\frac{\bar{\mu}
    c}{2}\hat{V}\cos\frac{\bar{\mu}
    c}{2}\nonumber\\
    &\quad\quad\quad\quad-\cos\frac{\bar{\mu}
    c}{2}\hat{V}\sin\frac{\bar{\mu}
    c}{2}\Big)\Big]\sin\frac{\bar{\mu}
    c}{2}\nonumber\\
&=:\sin\frac{\bar{\mu}
    c}{2}\,4\hat{A}\,\sin\frac{\bar{\mu}
    c}{2}.
\end{align}
It is easy to show that $\hat{A}$ is well defined and $|v\rangle$
is an eigenvector of $\hat{A}$. Furthermore, the eigenvalues of
$\hat{A}$ are real and negative. So $\hat{A}$ is a negative
definite self-adjoint operator on ${\cal
H}^{\mathrm{grav}}_{\mathrm{kin}}$. Hence,
$\hat{C}^{\mathrm{R}}_{\mathrm{grav}}$ is a negative-definite
self-adjoint operator on ${\cal
H}^{\mathrm{grav}}_{\mathrm{kin}}$. The action of
$\hat{C}^{\mathrm{R}}_{\mathrm{grav}}$ on the basis $|v\rangle$ of
${\cal H}^{\mathrm{grav}}_{\mathrm{kin}}$ is given by
\begin{align}
\label{hatC}
\hat{C}^{\mathrm{R}}_{\mathrm{grav}}|v\rangle=f'_+(v)|v+2\rangle+f'_o(v)|v\rangle+f'_-(v)|v-2\rangle,
\end{align}
where
\begin{align}
f'_+(v)&=\frac{27}{4}\sqrt{\frac{8\pi}{6}}\frac{K\ell_p}{\gamma^{3/2}}(v+1)\big(|v+2|-|v|\big),\nonumber\\
f'_-(v)&=f'_+(v-2),\quad f'_o(v)=-f'_+(v)-f'_-(v).
\end{align}
Thus, $\hat{C}^{\mathrm{R}}_{\mathrm{grav}}$ is again a difference
operator. Recall that by contrast to Eq. (\ref{Cr}), the
Hamiltonian constraint operator defined in \cite{aps} reads
\begin{align}
\label{Cs} \hat{C}_{\mathrm{grav}}&=\sin(\bar{\mu}
    c)\Big[\frac{3i\,\mathrm{sgn}(v)}{\pi\gamma^3
    \bar{\mu}^3\ell_p^2
    }\Big(\sin\frac{\bar{\mu}
    c}{2}\hat{V}\cos\frac{\bar{\mu}
    c}{2}\nonumber\\
    &\quad\quad\quad\quad-\cos\frac{\bar{\mu}
    c}{2}\hat{V}\sin\frac{\bar{\mu}
    c}{2}\Big)\Big]\sin(\bar{\mu}
    c)\nonumber\\
&=:\sin(\bar{\mu}
    c)\hat{A}\sin(\bar{\mu}
    c).
\end{align}
This shows a quantization ambiguity arising from the quantization
of the field strength of the gravitational connection.

To identify a dynamical matter field as an internal clock, we take
a massless scalar field $\phi$ with Hamiltonian $C_{\phi} =
|p|^{-3/2}\, p_\phi^2/2$, where $p_\phi$ denotes the momentum of
$\phi$. In the standard Schr\"{o}dinger representation, the matter
part of the quantum Hamiltonian constraint reads $
 \hat{C_{\phi}} =
\widehat{|p|^{-3/2}}\, \widehat{p_\phi^2}/2 $. Thus we get the
total constraint as $
 \hat{C}^{\mathrm{R}}=\frac{1}{16\pi G}\hat{C}^{\mathrm{R}}_{\rm
grav}+\hat{C}_\phi $.

\section{The classical limit and the modified Friedmann equation}

It has been shown in \cite{ta,dmy} that the improved Hamiltonian
constraint operator constructed in \cite{aps} has the correct
classical limit. In this section, we will show that the alternative
Hamiltonian constraint operator constructed in last section also has
the correct classical limit. Moreover, the effective Hamiltonian
incorporating higher order quantum corrections can also be obtained.
In order to do the semiclassical analysis, it is convenient to
introduce new conjugate variables by a canonical transformation of
$(c,p)$ as
\begin{align}
b:=\frac{\sqrt{\Delta}}{2}\frac{c}{\sqrt{|p|}}
\quad\mathrm{and}\quad
v:=\frac{\mathrm{sgn}(p)|p|^{3/2}}{2\pi\gamma\ell_p^2\sqrt{\Delta}},
\end{align}
with the Poisson bracket $\{b,v\}=1/\hbar$. In terms of these new
variables, the classical Hamiltonian constraint can be written as
\begin{align}
\label{Cbv} C&=\frac{C_{\rm grav}}{16\pi G}+C_\phi\nonumber\\
&=-\frac{27}{8\pi
G}\sqrt{\frac{8\pi}{6}}\frac{K\ell_p}{\gamma^{3/2}}\,b^2|v|+\frac{1}{2}\left(\frac{6}{8\pi\gamma\ell_p^2}\right)^{3/2}
\frac{K}{|v|}\,p_\phi^2.
\end{align}
Let us first consider the gravitational part. Since there are
uncountable basis vectors, the natural Gaussian semiclassical
states live in the algebraic dual space of some dense set in
${\cal H}^{\rm{grav}}_{\rm{kin}}$. A semiclassical state
$(\Psi_{(b_o,v_o)}|$ peaked at a point $(b_o,v_o)$ of the
gravitational classical phase space reads:
\begin{align}
(\Psi_{(b_o,\, v_o)}|=\sum_{v\in
\mathbb{R}}e^{-[(v-v_o)^2/2d^2]}e^{i\,b_o(v-v_o)}(v|,\label{coh}
\end{align}
where $d$ is the characteristic ``width'' of the coherent state.
For practical calculations, we use the shadow of the semiclassical
state $(\Psi_{(b_o,v_o)}|$ on the regular lattice with spacing 1
\cite{shad}, which is given by
\begin{align}
|\Psi\rangle \,=\, \sum_{n\in\Z}\,\left[e^{-(\epsilon^2/2)(n-N)^2}\,
\,e^{-i\, (n-N)b_o}\right] |n+\lambda\rangle ,\label{shad}
\end{align}
where $\lambda\in[0,1)$, $\epsilon = 1/d$ and we choose
$v_o=N+\lambda$, here $N\in\mathbb{Z}$. Since we consider large
volumes and late times, the relative quantum fluctuations in the
volume of the universe must be very small. Therefore we have the
restrictions: $ {1}/{N}\ll\epsilon\ll1$ and $ b_o\ll 1 $. One can
check that the state (\ref{coh}) is sharply peaked at $(b_o,v_o)$
and the fluctuations are within specified tolerance \cite{dmy,ta}.
The semiclassical state of matter part is given by the standard
coherent state
\begin{align}
(\Psi_{(\phi_o,p_{\phi})}|=\int
{\mathrm{d}}\phi\,e^{-[(\phi-\phi_o)^2/2\sigma^2]}\,e^{{i
p_\phi}(\phi-\phi_o)/\hbar}(\phi|\label{matterstate},
\end{align}
where $\sigma$ is the width of the Gaussian. Thus the whole
semiclassical state reads $(\Psi_{(b_o,\,
v_o)}|\bigotimes(\Psi_{(\phi_o,p_{\phi})}|$.

The task is to use this semiclassical state to calculate the
expectation value of the Hamiltonian operator to a certain accuracy.
In the calculation of
$\langle\hat{C}^{\mathrm{R}}_{\mathrm{grav}}\rangle$, one gets the
expression with the absolute values, which is not analytical. To
overcome the difficulty we separate the expression into a sum of two
terms: one is analytical and hence can be calculated
straightforwardly, while the other becomes exponentially decayed
out. We thus obtain (see the Appendix for details)
\begin{align}
\label{Cr1} \langle{
\hat{C}^{\mathrm{R}}}_{\mathrm{grav}}\rangle=&-54\sqrt{\frac{8\pi}{6}}\frac{K\ell_p}{\gamma^{3/2}}|v_o|
\Big[e^{-\epsilon^2}\sin^2b_o+\frac{1}{2}\left(1-e^{-\epsilon^2}\right)\Big]\nonumber\\
&\quad+O(e^{-N^2\epsilon^2}).
\end{align}
In the calculation of $\langle\hat{C}_{\phi}\rangle$, one has to
calculate the expectation value of the operator
$\widehat{|p|^{-3/2}}$. A straightforward calculation gives:
\begin{align}
\label{p32}
\langle\widehat{|p|^{-3/2}}\rangle=&\left(\frac{6}{8
\pi \gamma
\ell_p^2}\right)^{3/2}\,\frac{K}{|v_o|}\Big[1+\frac{1}{2|v_o|^2\epsilon^2}+\frac{5}{9|v_o|^2}\nonumber\\
&+O(1/|v_o|^4\epsilon^4)\Big]+O\big(e^{-N^2\epsilon^2}\big)
+O\big(e^{-\pi^2/\epsilon^2}\big).
\end{align}
Collecting these results we can express the expectation value of the
total Hamiltonian constraint, up to corrections of order
$1/|v_o|^4\epsilon^4$ and $e^{-\pi^2/\epsilon^2}$, as follows:
\begin{align}
\langle{ \hat{C}^{\mathrm{R}}}\rangle=&-\frac{27}{8\pi
G}\sqrt{\frac{8\pi}{6}}\frac{K\ell_p}{\gamma^{3/2}}|v_o|
\Big[e^{-\epsilon^2}\sin^2b_o\nonumber\\
&+\frac{1}{2}\left(1-e^{-\epsilon^2}\right)\Big]
+\frac{1}{2}\left(\frac{6}{8 \pi \gamma
\ell_p^2}\right)^{3/2}\frac{K}{|v_o|}\nonumber\\
&\times\left(p^2_{\phi}+\frac{\hbar^2}{2\sigma^2}\right)\left(1+\frac{1}{2|v_o|^2\epsilon^2}
+\frac{5}{9|v_o|^2}\right). \label{<C>}
\end{align}
Hence the classical constraint (\ref{Cbv}) is reproduced up to small
quantum corrections. Therefore, the new Hamiltonian operator is also
a viable quantization of the classical expression. For clarity, we
will suppress the label $o$ in the following. Using the expectation
value of the Hamiltonian operator in Eq. (\ref{<C>}), we can further
obtain an effective Hamiltonian with the relevant quantum
corrections of order $\epsilon^2,1/v^2\epsilon^2,\hbar^2/\sigma^2$
as
\begin{align}
{\cal H}^{\mathrm{R}}_{\mathrm{eff}}&=-\frac{27}{8\pi
G}\sqrt{\frac{8\pi}{6}}\frac{K\ell_p}{\gamma^{3/2}}\;|v|\left(\sin^2b+\frac{1}{2}\;\epsilon^2\right)\nonumber\\
&\quad\quad+\left(\frac{8\pi\gamma
\ell_p^2}{6}\right)^{3/2}\frac{|v|}{K}\;\rho\left(1+\frac{\hbar^2}{2\sigma^2p_\phi^2}+\frac{1}{2v^2\epsilon^2}\right),
\end{align}
where $\rho=\frac{1}{2}\left(\frac{6}{8 \pi \gamma
\ell_p^2}\right)^{3}\left(\frac{K}{v}\right)^2p_{\phi}^2 $ is the
density of the matter field. Then we obtain the Hamiltonian
evolution equation for $v$ by taking its Poisson bracket with
$\cal {H}^{\mathrm{R}}_{\mathrm{eff}}$ as
\begin{align}
\label{dotv} \dot{v}&=\{v,{\cal
H}^{\mathrm{R}}_{\mathrm{eff}}\}=\frac{27K}{2\sqrt{3\pi
G\hbar\gamma^3}}\;|v|\sin b\cos b.
\end{align}
Further, the vanishing of ${\cal H}^{\mathrm{R}}_{\mathrm{eff}}$
implies
\begin{align}
\label{sinb}
\sin^2b&=\frac{\rho}{\rho'_c}\left(1+\frac{\hbar^2}{2\sigma^2p_\phi^2}+\frac{1}{2v^2\epsilon^2}\right)
-\frac{\epsilon^2}{2},
\end{align}
where $\rho'_c=3/(2\pi G\gamma^2\Delta)$. The modified Friedmann
equation can then be derived from Eqs. (\ref{dotv}) and
(\ref{sinb}) as
\begin{align}
\label{rHub}
{H}^2&=\left(\frac{\dot{v}}{3v}\right)^2=\frac{8\pi G}{3}\,\rho'_c\sin^2b\cos^2b\nonumber\\
&=\frac{8\pi
G}{3}\,\rho\Big[1-\frac{\rho}{\rho'_c}\left(1+\frac{\hbar^2}{\sigma^2p_\phi^2}+\frac{1}{v^2\epsilon^2}\right)\nonumber\\
&\quad\quad\quad\quad+\frac{\hbar^2}{2\sigma^2p_\phi^2}+\frac{1}{2v^2\epsilon^2}-\frac{\epsilon^2}{2}
\frac{\rho'_c}{\rho}\Big].
\end{align}
Recall that, up to the quantum fluctuation of matter field, the
modified Friedmann equation given in \cite{dmy} reads
\begin{align}
\label{sHub} H^2&=\frac{8\pi
G}{3}\,\rho\left[1-\frac{\rho}{\rho_c}\left(1+\frac{1}{v^2\epsilon^2}\right)+\frac{1}{2v^2\epsilon^2}
-2\epsilon^2\frac{\rho_c}{\rho}\right],
\end{align}
where $\rho_c=3/(8\pi G\gamma^2\Delta)$. Comparing Eq. (\ref{rHub})
with (\ref{sHub}), we find that the leading order critical energy
density reads $\rho'_{\mathrm{crit}}=4\rho_{\mathrm{crit}}$. We can
also express the new modified Friedmann equation by using $\rho_c$
as
\begin{align}
\label{rHub2} H^2&=\frac{8\pi
G}{3}\,\rho\Big[1-\frac{\rho}{4\rho_c}\left(1+\frac{\hbar^2}{\sigma^2p_\phi^2}+\frac{1}{v^2\epsilon^2}\right)\nonumber\\
&\quad\quad\quad\quad\quad+\frac{\hbar^2}{2\sigma^2p_\phi^2}+\frac{1}{2v^2\epsilon^2}-2\epsilon^2
\frac{\rho_c}{\rho}\Big].
\end{align}

\section{Discussion}

\begin{figure}[!htb]
    \includegraphics[width=0.5\textwidth,angle=0]{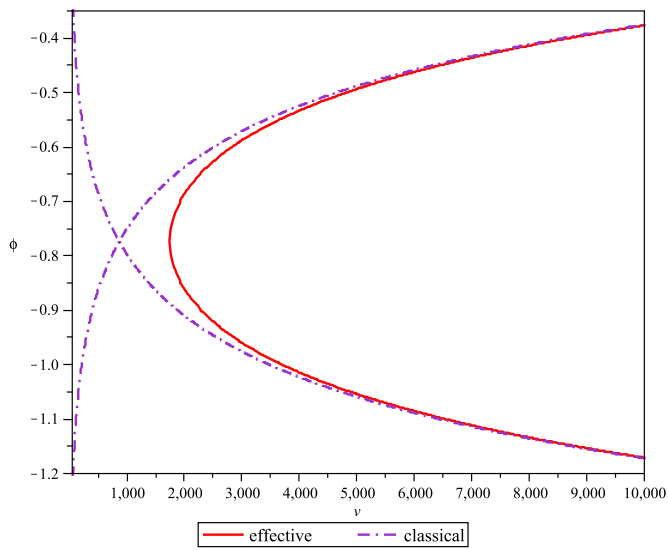}
    \caption{The effective dynamics represented by the observable
    $v|_{\phi}$ are compared to classical trajectories. In this
    simulation, the parameters were:
    $G=\hbar=1\,$, $p_{\phi}=10\,000,$
    $\epsilon=0.001\,,\sigma=0.01$ with initial data
    $v_o=100\,000\,$.}
    \label{fig:bounce}
\end{figure}

Because the properties of the alternative Hamiltonian constraint
operator in Eq. (\ref{Cr}) are similar to the one in Eq. (\ref{Cs}),
the physical Hilbert space, Dirac observables and so on investigated
in \cite{aps} can also be straight-forwardly obtained. Although
there are quantitative differences between the two versions of
quantum dynamics, qualitatively they have the same dynamical
features. In the leading order approximation, the universe would
bounce again from the contracting branch to the expanding branch
when the energy density of scalar field reaches to the critical
$\rho'_{\mathrm{crit}}=4\rho_{\mathrm{crit}}$. The quantum bounce
implied by Eq. (\ref{rHub}) is shown in Fig. \ref{fig:bounce}.

\begin{figure}[!htb]
    \includegraphics[width=0.5\textwidth,angle=0]{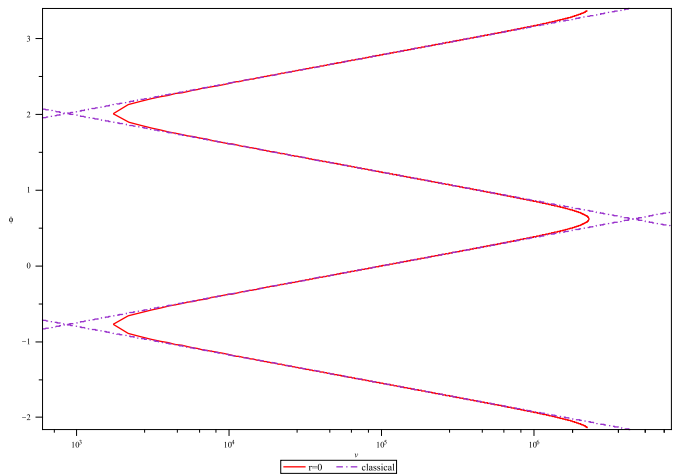}
    \caption{The cyclic model is compared with expanding and contracting classical trajectories.
    In this
    simulation, the parameters were: $G=\hbar=1$\,, $p_{\phi}=10\,000\,,\epsilon=0.001,\sigma=0.01$ with initial data
    $v_o=100\,000$.}
    \label{fig:recollapse}
\end{figure}

On the other hand, the key results discussed in \cite{dmy} can be
carried out similarly. It is easy to see from Eq. (\ref{rHub}) that,
while the term containing the quantum fluctuations of matter field
is qualitatively negligible, the asymptotic behavior of the quantum
geometric fluctuations plays a key role for the fate of the
universe. By the ansatz $\epsilon=\alpha(r)v^{-r(\phi)}$ with $0\leq
r(\phi)\leq1$, there are great possibilities for the expanding
universe to undergo a recollape in the future. The recollape can
happen provided $0\leq r<1$ in the large scale limit. Suppose that
the semiclassicality of our coherent state is maintained
asymtotically. This means that the quantum fluctuation $1/\epsilon$
of $v$ cannot increase as $v$ unboundedly as $v$ approaches
infinity. Thus the recollape is in all probability as viewed from
the parameter space of $r(\phi)$. For example, in the scenario when
$r=0$ asymtotically, besides the quantum bounce when the matter
density $\rho$ increases to the Planck scale, the universe would
also undergo a recollapse when $\rho$ decreases to $\rho_{\rm
coll}\approx\epsilon^2\rho'_c/2$. Therefore, the quantum
fluctuations again lead to a cyclic universe in this case. The
cyclic universe in this effective scenario is illustrated in Fig.
\ref{fig:recollapse}.
This is an amazing possibility that quantum gravity manifests
herself in the large scale cosmology. Nevertheless, the condition
that the semiclassicality is maintained in the large scale limit has
not been confirmed. Hence further numerical and analytic
investigations to the properties of dynamical semiclassical states
in the model are still desirable. It should be noted that in some
simplified completely solvable models of LQC (see \cite{bojow-r} and
\cite{acs}), the dynamical coherent states could be obtained, where
$r(\phi)$ approaches $1$ in the large scale limit. While those
treatments lead to the quantum dynamics different from ours, they
raise caveats to the inferred re-collapse.

In conclusion, the key features of LQC in this model, that the big
bang singularity is replaced by a quantum bounce and there are
great possibilities for an expanding universe to recollapse, are
robust against the quantization ambiguity which we have
considered.

\section*{ACKNOWLEDGMENTS}

This work is a part of project 10675019
supported by NSFC.

\section*{APPENDIX}

Let us calculate the expectation value of $\hat{C}^{\mathrm{R}}_{\rm
grav}$,
\begin{align}
\langle\hat{C}^{\mathrm{R}}_{\rm
grav}\rangle=\frac{(\Psi|\hat{C}^{\mathrm{R}}_{\rm
grav}|\Psi\rangle}{\langle\Psi|\Psi\rangle}.\label{expect}
\end{align}
Applying the Poisson resummation formula
\begin{align}
\sum_{n=-\infty}^{\infty}g(n+x)=\sum_{n=-\infty}^{\infty}e^{2\pi
in x}\int_{-\infty}^{\infty}g(y)e^{-2\pi iny}\mathrm{d}y
\label{poinsum}
\end{align}
to the norm of the shadow state
(\ref{shad}), one obtains
\begin{align}
&\langle\Psi|\Psi\rangle=\sum_{n\in\Z}e^{-\epsilon^2n^2}=\frac{\sqrt{\pi}}{\epsilon}\left(1+
O(e^{-\frac{\pi^2}{\epsilon^2}})\right).
\end{align}
By Eq. (\ref{hatC}), we obtain the action of the gravitational
Hamiltonian operator on the shadow state as
\begin{align}
\hat{C}^{\mathrm{R}}_{\mathrm{grav}}|\Psi\rangle=&\sum_{n\in\Z}\,
e^{-(\epsilon^2/2)(n-N)^2}
 e^{-i(n-N)b_o}\nonumber\\
&\times\Big[f'_+(n+\lambda)|n+\lambda+2\rangle+f'_o(n+\lambda)|n+\lambda\rangle\nonumber\\
&\quad\quad+f'_-(n+\lambda)|n+\lambda-2\rangle\Big].\nonumber
\end{align}
Then a straightforward calculation shows that
\begin{align}
&(\Psi|\hat{C}^{\mathrm{R}}_{\mathrm{grav}}|\Psi\rangle\nonumber\\
 &=\sum_{n\in\Z}e^{-\epsilon^2(n-N)^2}\Big[2e^{-\epsilon^2}\cos(2b_o)f'_+(n+\lambda-1)\nonumber\\
 &\quad\quad\quad\quad\quad\quad\quad\quad -(f'_+(n+\lambda)+f'_-(n+\lambda))\Big]\nonumber\\
 &=\frac{27}{4}\sqrt{\frac{8\pi}{6}}\frac{K\ell_p}{\gamma^{3/2}}\left[2e^{-\epsilon^2}\cos(2b_o)\bar{S}_0
 -(\bar{S}_1+\bar{S}_{-1})\right],
 \end{align}
where
\begin{align}
&f'_+(n)=\frac{27}{4}\sqrt{\frac{8\pi}{6}}\frac{K\ell_p}{\gamma^{3/2}}(n+1)\big(|n+2|-|n|\big),\nonumber\\
&f'_-(n)=\frac{27}{4}\sqrt{\frac{8\pi}{6}}\frac{K\ell_p}{\gamma^{3/2}}(n-1)\big(|n|-|n-2|\big),\nonumber
\end{align}
and we have set
\begin{align}
\bar{S}_{m}:=&\sum_{n\in\Z}e^{-\epsilon^2(n-N)^2}(n+\lambda+m)\nonumber\\
 &\quad\times\big(|n+\lambda+m+1|-|n+\lambda+m-1|\big),\nonumber\\
S_{m}:=&2\sum_{n\in\Z}e^{-\epsilon^2(n-N)^2}(n+\lambda+m)\nonumber\\
 =&2(N+\lambda+m)\sum_{n\in\Z}e^{-\epsilon^2n^2}.\nonumber
\end{align}
We may bound $|\bar{S}_{m,k}-S_{m}|$ by an exponentially
suppressed term:
\begin{align}
|\bar{S}_{m}-S_{m}|=&\,\bigg|\sum_{n\in\Z}e^{-\epsilon^2(n-N)^2}(n+\lambda+m)\nonumber\\
&\quad\quad\times\big(|n+\lambda+m+1|\nonumber\\
&\quad\quad\quad-|n+\lambda+m-1|-2\big)\bigg|\nonumber\\
=&\,\bigg|\sum_{-1<n+\lambda+m<1}e^{-\epsilon^2(n-N)^2}(n+\lambda+m)\nonumber\\
&\quad\quad\times\big(2(n+\lambda+m)-2\big)\nonumber\\
&\quad\quad+\,4\sum_{n+\lambda+m\leq-1}e^{-\epsilon^2(n-N)^2}(n+\lambda+m)\bigg|\nonumber\\
\leq &\,\bigg|\sum^{0}_{n+m=-1}e^{-\epsilon^2(n-N)^2}(n+\lambda+m)\nonumber\\
&\quad\quad\times\big(2(n+\lambda+m)-2\big)\bigg|\nonumber\\
&\quad\quad+\,4\bigg|\sum_{n+m\leq-1}e^{-\epsilon^2(n-N)^2}(n+m)\bigg|\nonumber\\
&\quad\quad+4\lambda\sum_{n+m\leq-1}e^{-\epsilon^2(n-N)^2}\nonumber\\
=&\bigg|e^{-\epsilon^2(m+N)^2}\lambda(\lambda-1)\nonumber\\
&\,+e^{-\epsilon^2(m+N)^2}(\lambda-1)(\lambda-2)\bigg|\nonumber\\
&+\,4\sum_{n\geq1}e^{-\epsilon^2(n+m+N)^2}n\nonumber\\
&+4\lambda\sum_{n\geq1}e^{-\epsilon^2(n+m+N)^2}
\end{align}
Using the Euler-Maclaurin summation, for some positive constant
$\alpha$ independent of $N,m$ or $\epsilon$, we obtain:
\begin{align}
4\sum_{n\geq1}e^{-\epsilon^2(n+m+N)^2}n\leq &\alpha
e^{-N^2\epsilon^2}+4\int_{1}^{\infty}e^{-\epsilon^2(x+m+N)^2}x{\mathrm{d}}x\nonumber\\
<&\alpha
e^{-N^2\epsilon^2}+\frac{2}{\epsilon^3(m+N)}e^{-\epsilon^2(m+N)^2}.
\end{align}
Hence we obtain $|\bar{S}_{m}-S_{m}|=O(e^{-\epsilon^2N^2})$.
Therefore, we have
\begin{align}
(\Psi|\hat{C}^{\mathrm{R}}_{\mathrm{grav}}|\Psi\rangle&=\frac{27}{4}\sqrt{\frac{8\pi}{6}}\frac{K\ell_p}{\gamma^{3/2}}
\Big[2e^{-\epsilon^2}\cos(2b_o)S_0\nonumber\\
 &\quad\quad-(S_1+S_{-1})\Big]+O(e^{-N^2\epsilon^2}).
 \end{align}
Finally, we can collect the above terms to calculate the expectation
value (\ref{expect}). Using
$\langle\Psi|\Psi\rangle=\sum_{n\in\Z}e^{-\epsilon^2n^2}$, we have
\begin{align}
\langle\hat{C}^{\mathrm{R}}_{\mathrm{grav}}\rangle
&=27\sqrt{\frac{8\pi}{6}}\frac{K\ell_p}{\gamma^{3/2}}(N+\lambda)\left[e^{-\epsilon^2}\cos(2b_o)-1\right]\nonumber\\
&\quad+O(e^{-N^2\epsilon^2})\nonumber\\
&=-54\sqrt{\frac{8\pi}{6}}\frac{K\ell_p}{\gamma^{3/2}}|v_o|\left[e^{-\epsilon^2}\sin^2b_o
+\frac{1}{2}(1-e^{-\epsilon^2})\right]\nonumber\\
&\quad\quad+O(e^{-N^2\epsilon^2}).
\end{align}

\end{document}